# ASimLab 1.1, Software for Sensitivity and Uncertainty Analysis, tool for sound modelling

N.Giglioli, A.Saltelli,
Joint Research Centre
European Commission
Institute for Systems Informatics and Safety
Ispra, Italy
nicla.giglioli@jrc.it
andrea.saltelli@jrc.it

**Abstract**

The aim of this paper is to present and describe SimLab 1.1 (Simulation Laboratory for Uncertainty and Sensitivity Analysis) software designed for Monte Carlo (MC) analysis that is based on performing multiple model evaluations with probabilistically selected model input. The results of these evaluations are used to determine both the uncertainty in model predictions and the input variables that drive this uncertainty. This methodology is essential in situations where a decision has to be taken based on the model results; typical examples include risk and emergency management systems, financial analysis and many others. It is also highly recommended as part of model validation, even where the models are used for diagnostic purposes, as an element of sound model building. SimLab allows an exploration of the space of possible alternative model assumptions and structure on the prediction of the model, thereby testing both the quality of the model and the robustness of the model based inference.

## 1. Introduction

According to Hornberger and Spear (1981) "… most simulation models will be complex, with many parameters, state-variables and non linear relations. Under the best circumstances, such models have many degrees of freedom and, with judicious fiddling, can be made to produce virtually any desired behaviour, often with both plausible structure and parameter values. "

The problem highlighted by Hornberger is acutely felt in the critical segment of the modelling community. An economist, Edward E. Leamer, suggests the following:

"I have proposed a form of organised sensitivity analysis that I call "global sensitivity analysis" in which a neighbourhood of alternative assumptions is selected and the corresponding interval of inferences is identified. Conclusions are judged to be sturdy only if the neighbourhood of assumptions is wide enough to be credible and the corresponding interval of inferences is narrow enough to be useful."

This awareness of the dangers implicit in selecting a model structure as *true* and working happily thereafter leads naturally to the attempt to rigorously map alternative model structures or working hypotheses into the space of the model predictions.

Tools like SimLab favour such an attempt. It aims to provide the modeller with a general driver to perform MC-type analysis where both uncertainty in data and parameters, and that arising from higher order uncertainties can be accommodated.

The natural extension of an analysis aimed at quantifying the impact of different sources of uncertainty into the output, is the analysis of *how much* each source of uncertainty weights on the model prediction. One of the possible ways to quantify the importance of the input factor with respect to the model output is to apply global quantitative sensitivity analysis methods. Here the expression "Global Sensitivity Analysis" takes on an additional meaning, with respect of that proposed by Leamer, in that a decomposition of the total model uncertainty is sought.

SimLab software is a tool that helps the analyst to take into account the effect of the input uncertainties that affect the model output. It is designed to explore the space of the input uncertainties, via either MC or other strategies, and then it uses the results to determine both uncertainty in model predictions and how these are apportioned to the input factors (Saltelli et al. 2000a). The analysis involves five steps: the selection of ranges and distributions for each factors, generation of a sample from the ranges and distributions specified in the first step; evaluation of the model for each element of the sample; uncertainty analysis and sensitivity analysis.

An important feature of the software is its flexibility with respect to the model to be analysed. SimLab strives to facilitate as much as possible the troublesome task of interfacing a given model, in most instances conceived to run on a single set of input values at a time, with the MC driver that allows multiple execution of the model. Processing the model output, SimLab performs then Uncertainty Analysis (UA) and Sensitivity Analysis (SA).

This paper is organised as follows. Section 2 is devoted to definitions and description of the methodologies for performing UA and SA with examples of the applications. In section 3, capabilities and functionality of the software are presented. Section 4 discusses more in detail examples of applications, to illustrate the full potentiality of the proposed software and the underlying methodology.

## 2. What is UA and SA

2.1 Definitions

There are different definitions related to SA depending on the goal of the analysis and on the point of view of the analyst.

For an engineer, it could be crucial to test the model reliability under different assumptions of lifetime, fault tree structure and the like. For a statistician it could be interesting to test the robustness of a statistical model with respect to distributional assumptions. In econometrics it is important to test the stability of a regression model with respect to all factors excluded from it. For a modeller involved with geo-referenced analysis, it is crucial which resolution level to use, and so on.

Possibly as a result of this myriad of possible settings, many different meaning are attached to SA in the various communities. In general, UA determines the uncertainty in model predictions due to imprecisely known of input variables, while SA determines the contribution of individual input variable to the uncertainty in model predictions (see Helton, 1993). For the purpose of describing SimLab, we shall say that SA studies the relationship between the information flowing in and out of the model (Saltelli et al. 2000a). An alternative definition for the task carried out by SimLab is *importance assessment*. As it will be seen in the examples, a crucial element of the analysis is the identification of the output variable of interest.

2.2 Characteristics of the methods

In SimLab it is possible to choose among different sampling methods like, random, fixed sampled, Latin Hypercube, Morris, Replicated Latin Hypercube, Quasi Random $LP_\tau$ (read: LP-tau) and FAST (Fourier Amplitude Sensitivity Test).

The choice of the method is a delicate step as it conditions the SA method to be used (for discussion see Campolongo et al. 2000a). For this reason, the software contains a quick help for the non-expert user.

The methods implemented are described in brief in the next subsections.

*2.2.1 Screening method: The Morris method*

We submit that one important goal in modelling is to identify the important factors, and to do this, the choice of a well-designed experiment is essential.

Screening methods are meant to deal with models containing hundreds of input factors, or with very computationally expensive models. They are economical from a computational point of view, but as a drawback, they tend to provide qualitative sensitivity measures, i.e. they rank the input factors in order of importance, but do not quantify how much a given factor is more important than another. There is clearly a trade-off between computational cost and information.

Several approaches to the problem of screening have been proposed in the literature. The one-at-a-time (OAT) Morris method has been implemented in SimLab. For a review on the method see Campolongo et al., 2000b.

*2.2.2 Regression Analysis and Correlation Measures*

Generally more expensive than screening tests, regression methods are quantitative but only if the model does not depart too much from linearity. Regression is possible in SimLab by selecting one of the following sampling methods: random, fixed sampled, Latin Hypercube and Quasi Random. The analysis yields the SRC (Standardized Regression Coefficient), the PCC (Partial Correlation Coefficient), PEAR (Pearson product moment correlation coefficient) and the coefficient of model determination $R_y^2$. The validity of the SRC, PEAR and PRCC as a measures of sensitivity is conditional on the degree to which the regression model fits the data (in this case, reproduce the original model), i.e. to $R_y^2$.

Rank transformed statistics (Standardised Rank Regression Coefficients (SRRC's) or the Partial Rank Correlation Coefficients (PRCC's) and SPEA (Spearman coefficient)) are also performed by default. They are useful when the model is not linear but is additive (i.e. $Y = \sum_i f_i(x_i)$, where $f_i$ is a function of $X_i$), and are associated with higher $R_y^2$. The increased explicative power of the rank based test comes at the cost of a bias in the analysis as discussed in Saltelli and Sobol', 1995.

*2.2.3 FAST, Extended FAST and Importance Measures*

FAST (Fourier Amplitude Sensitivity Test), Extended FAST and Importance Measures are variance based techniques (see Chan et al., 2000). In this case, the SA is based on estimating the fractional contribution of each input factor $X_i$ to the variance of the model output $Y$. In order to calculate the sensitivity indices for each factor, the total variance $V$ of the model output is decomposed as

$$V = \sum_i V_i + \sum_{i<j} V_{ij} + \sum_{i<j<m} V_{ijm} + ... + V_{12...k}$$

where $V_i = V(E(Y|X_i = x_i^*))$, and $V_{ij}$ is $V(E(Y|X_i = x_i^*, X_j = x_j^*)) - V_i - V_j$. $E(Y|X_i = x_i^*)$ denotes the expectation of $Y$ conditional on $X_i$ having a fixed value $x_i^*$. The decomposition of $V(Y)$ is unique if the $X_i$ are independent. The sensitivity index $S_i$ for the factor $X_i$ is defined as $V_i/V$. The reason for that is intuitive: if the inner mean $E(Y|X_i = x_i^*)$ varies considerably with the selection of a particular value $x_i^*$ for $X_i$, while all the effects of the $X_j$'s, $j \neq i$ are being averaged, then surely factor $X_i$ is an influential one. Estimation procedures for $S_i$ are FAST, the method of Sobol', and others (Chan et al., 2000). Importance measures (Iman and Hora, 1990) are also based on different estimates of the same quantity, $V(E(Y|X))$. This quantity can be shown to be equivalent to the Sobol' and FAST' first order sensitivity indices. For orthogonal inputs and linear models, $V(E(Y|X))$ is identical to the squared standardised regression coefficients (Chan et al. 2000).

Higher order sensitivity indices, responsible for interaction effects among factors, are rarely estimated in computational experiments, as in a model with $k$ factors the total number of indices (including the $S_i$'s) that should be estimated is as high as $2^k - 1$. This problem is sometimes referred to as the curse of dimensionality. However interactions may have a strong impact on the output uncertainty especially when $k$ is large and factors are varied on a wide scale, as often happens in numerical modelling. It is evident that interactions are of interest for the analysis, as synergetic effects involving more than one factor could, for instance, lead to extreme model outputs.

A method, which is able of accounting for interactions and simultaneously coping with the curse of dimensionality, is the extended FAST (Saltelli et al. 1999). The extended FAST can yield estimates of the total sensitivity indices. $S_{Ti}$ defined as the sum of all the indices ($S_i$ and higher orders) where a given factor $X_i$ is included. This concentrates in one single term all the effects involving $X_i$. For additive models with no interaction between the factors $X_i$, $S_i = S_{Ti}$ and $\sum_i S_i = 1$. The estimation of the total sensitivity indices $S_{Ti}$ makes the analysis affordable from a computational point of view, as we only need $k$ total indices for decomposing quantitatively the output variance V. Furthermore, the extended FAST allows the simultaneous evaluation of the first and total effect indices. The estimation of the pair $(S_i, S_{Ti})$ is important to appreciate the difference between the impact of factor $X_i$ alone on $Y$ (that is $S_i$) and the overall impact of factor $X_i$ through interactions with the others on $Y$ (provided by $S_{Ti}$).

Variance based methods such as Sobol' and the extended FAST display a number of attractive features for SA:
- Model independence: the sensitivity measure is model independent. It works for non-linear and non-additive models, unlike methods based on linear regression such as the standardised regression coefficients (Helton 1993).
- The measure captures the influence of the full range of variation of each factor.
- The measure captures interaction effects; this can be a crucial issue for a design problem, or for a risk analysis study.
- The methods can treat "sets" of factors as one single factor. This means that the analysis can be performed by partitioning the $k$ factors in a few subgroups and work on these rather than on the individual factors. The added value of performing by an analysis on groups of factors is clear: in complex models uncertain factors might pertain to different logical levels, and it might be desirable to decompose the uncertainty according to these levels.

An important aspect of the variance-based methods described in this section is the variance decomposition discussed above. When the inputs are correlated the decomposition loses its unicity and the concept of terms of higher order becomes troublesome (see Saltelli et al., 2000c).

In this case we can still compute the first order indices $V(E(Y|X))$, using Sobol' method and a rather more expensive computational algorithm (see details in SimLab 2000, Saltelli et al. 2000c).

*2.2.4 Remarks on the methods.*

SimLab does not compute local sensitivity analysis measures such as derivatives of the output with respect to some of the inputs. For these, specialised software are available (see et Grievank 2000, Turanyi 1990 and Turanyi et al. 2000). Also, SimLab does not allow mini-max or extreme bounds analysis (computing the model for extreme-value combination of the input, in order to identify extreme values of the output). We simply do not recommend such an approach (see also Leamer, 1991).

2.3 Fields of application

The analyses provided by SimLab can be seen as an element of model verification and validation. Monte Carlo model verification is based on analysing the output from different model parameters combinations. This allows checking whether the model output obtained is the expected one (e.g. this can be used to flag when the model produces physically implausible outcomes). On top of this, SA offers an additional layer of verification; in that suspect sensitivity of factors considered to be unimportant could be a symptom of coding or modelling error.

As discussed in the literature, model validation should be seen in the context of the model objectives e.g. procedures could differ between diagnostic and prognostic models (see Saltelli and Scott 1997).

SA could be used to identify critical regions in the space of input parameters, or to identify among a group of factors the few influential ones. In some applications it could be important to ascertain if a subset of factors may account for most of the total model output variance, thus allowing the non important factors to be fixed. In this case the choice of the model output of interest is crucial. One such case is discussed in section 4 and concerns a GIS-based model reduction exercise in which a good assessment of the input factor is beneficial to data resolution management, as this is an expensive step of the model building. Another advantage is the possibility to reinforce model output predictions with their uncertainty.

SA has been employed in a scenario analysis (see Draper et al. 2000). In this situation, SA aims at finding relevant sources of uncertainty in a model that predicts what would happen if deep geosphere disposal barriers were compromised in the future by geological faulting, human intrusion and/or climatic change.

Level E model is another application where SA has been carried out (see Saltelli et al. 2000b). The model describes mass transfer with chemical reaction in a porous medium. The migrating species are radioactive, and the attention is devoted to the total dose to man resulting from all the radionuclides, once these reach the biosphere. A panel of experts has defined the input factors and distribution for the Level E model. A FAST analysis by groups has been performed to assess system performance obtaining a better understanding of the model behaviour. Level E model is a demo included in SimLab.

In the field of economic time series the attention is devoted to the estimation of unobserved movements like trend and seasonally adjusted series. These can be seen as conditional on a model specification and on the estimated parameter values. One aim for the econometrician could be to assess the uncertainty in the estimators of the components due to the uncertainty in the model specification and model parameters. Furthermore, SA techniques are used to isolate the sources of uncertainty that affect the trend, seasonal and short-term movements in time series (Planas et al., 2000).

SA methodologies could also be used in financial field to quantify market or credit risks. Market risk is associated with the unknown behaviour of the market variables while credit risk arises from the possibility of default by one of the counterparts involved in a financial transaction. It is essential for a proper functioning of a bank to find strategies to manage risk and SA techniques are very useful for this purpose (Campolongo et al., 2000c). For example, to build a risk-less portfolio it is essential to estimate the maximum loss that any given financial instrument in the portfolio might cause (UA) and what was this due to (SA). For a general discussion on other settings or reasons for SA, the reader is referred to (Saltelli and Scott, 1997).

## 3. Description of the software

SimLab runs under Windows/NT, and the memory required is at least 32 Mb. Its user-friendly menu allows different UA and SA strategies to be selected. An on-line help is also available to the user.
The examples included in SimLab as Demo are:
- a linear model
- the Ishigami function
- Sobol G function
- the Level E model.

The first three are simple analytic formulae, while the fourth is a model of medium complexity (solves a system of Partial Differential Equation). SimLab is designed for computing Monte Carlo Analysis that consists of five steps (better defined in Section 2) that are computed by the three modules: Pre-processor, Model specification and Post-processor modules (see Figure 1).

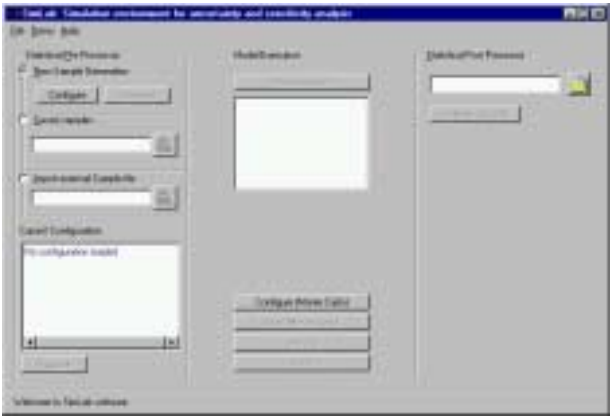

**Figure 1: SimLab. The three vertical panels correspond to the three modules**

3.1   Pre-processor module

The Pre-Processor module allows the user to define the list of factors that represents the input of the model, to specify the distribution/parameter assumptions on each factor and the correlation structure (see Figure 2).

A wide range of distributions is available as well two types of correlation structures: rank-correlation based (Iman and Conover (1982)) and "Tree correlation with copula" (Meeuwissen and Cooke (1994)).

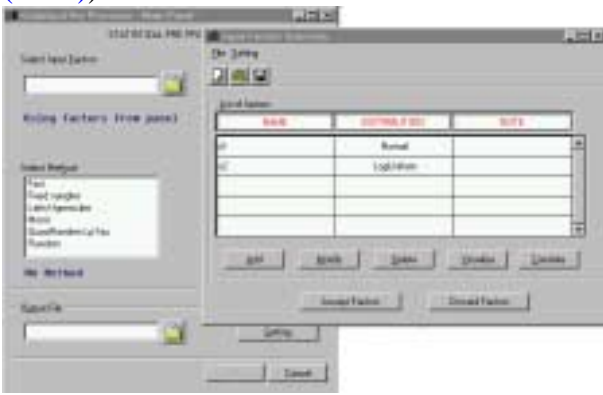

**Figure 2: Prep-processor module, list of factor and methods**

The next step is the sample generation according to the type of methodology the analyst wants to use. Different choices are available in SimLab: random sampling, quasi-random $LP_\tau$, Replicated Latin Hypercube, Latin Hypercube, classic and extended FAST (Fourier Amplitude Sensitivity Test), Morris and fixed sample. The last one is useful in the testing phase to run the system with known inputs.

When the sample is generated, for each factor, descriptive statistics and information related to correlation characteristics are estimated. The sampling distribution of each factor and the correlation characteristics amongst the factors can be visualised, by using histograms, cobweb plots (Cooke et al. 2000) or scatter plots respectively.

3.2   Model specification module

There are two possibilities of simulation here:
1. the user can link SimLab to his/her external model via executable files or statistical and mathematical packages;
2. the user can define a model within SimLab by using a simple equation editor (formula parser).

When the model is complex the first possibility is clearly the only one available. In the examples of Section 4 SimLab has been interfaced to external software (see Figure 5).

The external model performs model evaluation at each sample point, and yields a set of model outputs.

### 3.3 Post-processor module

The Post-processor module performs the final steps of the MC analysis, UA (see Figure 3) and SA (see Figure 4).

For the UA, this module provides statistics, confidence bounds, percentiles, histograms and cumulative plots of the model outputs. Non-parametric techniques are also implemented (i.e. Tchebycheff and Kolmogorov confidence bounds).

SA techniques implemented in this section are those described in 2.2.1 to 2.2.3, i.e. the regression, Morris and Variance based methods. For the extended FAST, the Post-processor module also incorporates the computation of sensitivity indices by groups; that is, factors are partitioned into groups according to (known) similar characteristics. When the input are correlated, $V(E(Y|X))$-based indices can be computed.

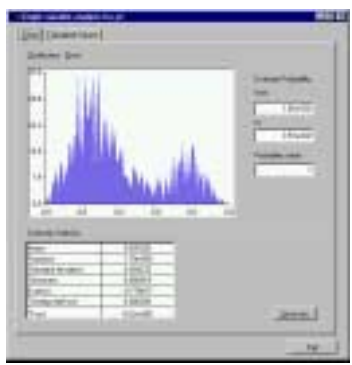

**Figure 3: Post processor module: an example of UA**

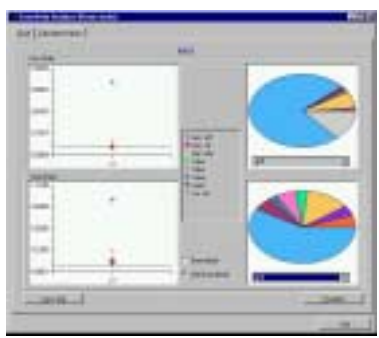

**Figure 4 Post processor module: an example of SA with FAST method**

## 4. Worked examples

Two applications are described here: in the first one, the model is an environmental index. In the second, the model is a hydrological model based on a geographically distributed system.

### 4.1 Environmental indices

The aim of this exercise is to compute a geographically distributed environmental index to be used for policy-making (Tarantola et al., 2000). Supposing that a new road has to be built and two options have been proposed, the one with lower environmental impact should to be chosen. Input data which are available at different resolution levels, are related to the quality of soil (drainage, acidity and depth), land cover maps (watercourses, protected area, urban or rural area) and roads network with planned variants proposed.

Each of the parcels in the area crossed by the options has been assigned three indicators as follows:

a) *environmental value loss*: given the specific problem at hand and the data available, for parcels outside protected areas a value of zero is assumed; the parcel is considered "lost" for the environment, with a weight range of [100; 200], if it is inside a protected area (see table 1).

| Uncertainties In Data Measurement | Factor $X_i$ | Distribution Assumed |
|---|---|---|
| Soil depth | $\varepsilon^{sd}$, Multiplicative error associated with the soil depth map | U(0.8,1.1) |
| Acidity | $\varepsilon^{pH}$, Multiplicative error associated with the acidity map. | U(0.8,1.1) |
| **Uncertainties In The Model Specifications** | | |
| Weights associated to the map of land use and to the map of protected areas | $W_{urban}$ $W_{agr}$ $W_{water}$ $W_{prot}$ | U(80,100) U(30,80) U(100,120) U(100,200) |
| Buffer size around existing and planned roads | $Width_{buffer}$ | U(30,50) |
| **Uncertainties In Data Selection** | | |
| To select between two resolutions | $X_{resolution}$ | U(0,1) |

**Table 1 List of input factors**

b) *economic costs*: three types of land cover are distinguished (urban and agricultural land use, water courses), with respective economic loss ranges of [80; 100], [30; 80], [100; 120] (see table 1). For urban and agricultural land use, the costs (weights) are calculated on the basis of land prices, while the "water course" value range reflects the high additional costs of building a highway over a river.

c) *potential for cultivation*: the indicator representing potentials for cultivation is computed from the soil characteristics (i.e., depth, drainage, and acidity) using a simplified version of a model available in the literature (Sanesi, 1982). For each parcel, a loss in the range of [30; 80] points is added, reflecting the fact that valuable soil is being sacrificed for a motorway. This cost reflects impacts on the competitiveness of agriculture and on the social network in agriculture-based rural societies that are not adequately captured by the pure land value.

The three indicators are added and for each option proposed, the index *I* is evaluated (i.e. $I_1$ and $I_2$).

The model output *Y* is then defined as $Y = \log_{10} \frac{I_1}{I_2}$, such that, if *Y* is positive (negative), the first (second) option is environmentally preferable. The magnitude of *Y* indicates how much a variant is better than the alternative one.

The model has been coded using Avenue, which is implemented in ArcView 3.1[1] and the generation of the input sample and the global sensitivity analysis have been performed using SimLab (SimLab, 2000).

Figure 5 shows how Simlab has been interfaced with ArcView. The sample file generated in Prep panel was fed as input in ArcView where the model evaluations are computed. These are saved in an output file analysed into Spop module.

---
[1] ArcView GIS Version 3.1, Copyright© Environment System Research Institute, Inc. All rights reserved

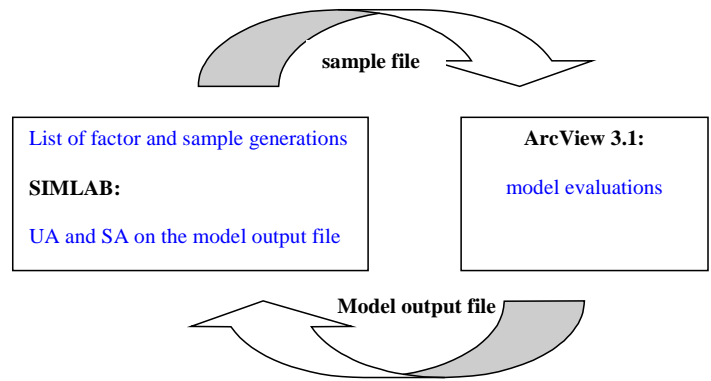

**Figure 5 Scheme of how SimLab could be linked to external software ArcView.**

As the inputs were considered to be uncorrelated, the extended FAST-based method was used. The full analysis required $Nk = 1096$ sample points and as many model evaluations. This has allowed FAST to compute the pair $(S_i, S_{Ti})$ for each of the eight factors.

UA (see figure 6) shows that, despite of poorly known data, subjective uncertainty and parametric uncertainty, the second road option is preferable to the first, given that 100% of the model results have positive values for *Y*.

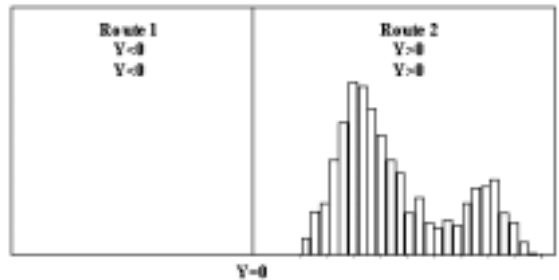

**Figure 6: Histograms of the model output *Y***

The SA (see table2) indicates that the bimodal nature of the histogram is due to the uncertainty in the selection of the resolution level (59% of the total output variance is accounted for by the factor itself) and to the uncertainty in the selection of the buffer width around the highway (23%). The sum of the first order indices, an indicator of model additivity, is 0.92. This value, close to one, indicates that the model is nearly additive. A small fraction (8%) of the output variance, i.e. $1 - \sum_i S_i = 0.08$, is due to interactions among the factors.

| $X_i$ | $S_i$ | $S_{Ti}$ |
|---|---|---|
| $X_{resolution}$ | 0.59 | 0.78 |
| $Width_{buffer}$ | 0.23 | 0.45 |
| $\varepsilon^{sd}\ \varepsilon^{pH}\ W_{urban}\ W_{agr}\ W_{water}\ W_{prot}$ | (0 - 0.02) | (0.04 - 0.14) |

**Table 2: Sensitivity Indices**

The influence of the other factors on the model output uncertainty is almost negligible, given the small values of their total effect indices (see Table 2). This indicates, that it would be unwise --say, to try to narrow the ranges in the assignment of the weights of by making further surveys among stakeholders. This would not improve the quality of the decision process.

4.2 Hydrological risk example

In this example, the risk of flooding of a small alpine valley is considered (see Crosetto et al., 2000). Risk assessors and decision-makers can make use of a model as part of the decision process:
- to decide on licensing a recreational area in the region
- as part of an early warning system, to decide whether to evacuate population

In order to effectively support the process, such predictions must be provided with their characterised uncertainty, given the critical nature of the decisions. UA is performed to test whether or not model predictions remain within some desired target bounds. The use of global SA helps to identify the quality requirements for geo-referenced input data that guarantee satisfactory support to predicting flooding risk, and to test the quality of the model.

The model CASC2D, developed in the GRASS[2] software, has been used to generate the model output which have been analysed using SimLab (SimLab, 2000). The main cartographic and hydrologic data used in the simulation are a DEM (digital elevation model), a land cover map; a lithology map; a vector data (road, hydrologic network, etc.) and finally a time series of rainfall intensity. The uncertainty of each of these input factors has been characterised (Crosetto et al., 2000). Since the number of input factors is quite high, a preliminary analysis, based on the Morris technique, was performed to screen out the important factors (56 model evaluations). The Extended FAST was applied on these factors and 485 model evaluations were performed. The output of interest is the river level H, for which the following values were obtained:

Mean(H) = 3.23 m
Standard deviation(H) = 0.49 m

| Input factors | $S_i$ | $S_{Ti}$ |
|---|---|---|
| rainfall intensity | 0.80 | 0.84 |
| vector data | 0.05 | 0.08 |
| Interception Map | 0.03 | 0.06 |
| initial soil moisture values | 0.02 | 0.05 |
| porosity values | 0.02 | 0.04 |

**Table 3: Sensitivity Indices**

The first order ($S_i$) and total sensitivity indices ($S_{Ti}$) computed using the extended FAST technique are reported in table 3. It can be noted that $S_i$ and $S_{Ti}$ are very similar, indicating that none of the factors is involved in significant interactions. The contribution of interactions among factors within this model is 10%, as quantified by $1 - \sum_i S_i$. It is clear that major source of uncertainty is represented by the rainfall ($S_{Ti} = 0.84$). The only way to significantly reduce the prediction uncertainty would be to improve the precision in forecasting this input factor, which is hardly surprising in this problem setting.

SA in this case reassures us that the quality of the information collected is sufficient for the task at hand (e.g. we do not need a finer resolution for any of the non-rainfall factors). Clearly this statement is not of general validity given the model but only for the model given this particular task (flooding risk prediction at valley level).

## 5. Conclusion

Model input data are subject to sources of uncertainty including errors of measurement, inadequate sampling resolution, etc. Furthermore, the model itself can include conceptual uncertainty, i.e. uncertainty in model structures assumptions and specifications. All this imposes a limit on confidence in model output. SimLab contributes to good modelling practice by providing a measure of the above. The test cases described give a hint of how SimLab can be useful in model based decision making. As discussed in the introduction, this is but one of the possible settings where quantitative UA and SA are of use, in fields where models are employed.

*Acknowledgements*

SimLab has obtained grants from the European Commission DG XIII-D (SUP COM projects in 1997 and 1998, Validation and Integration, Demonstration and Marketing of a software for Uncertainty &

---

[2] For more detailed information about GRASS see: www.baylor.edu/~grass.

Sensitivity Analysis). The test case on GIS has been partially funded by the Statistical Office of the European Union (EUROSTAT, SUPCOM 98).